\documentclass{kluwer}
\usepackage{graphics}
\usepackage{amssymb}

\begin{document}
\begin{article}
\begin{opening}
\title{\centering \normalsize \bf RELATIVE TIMING OF SOLAR FLARES OBSERVED AT DIFFERENT
WAVELENGTHS}
\runningtitle{RELATIVE TIMING OF SOLAR FLARES}
\author{A.~Veronig}
\runningauthor{A. VERONIG ET AL.}
\institute{Institute for Geophysics, Astrophysics and Meteorology,
University of Graz, Universit\"atsplatz 5, A-8010 Graz, Austria}
\author{B.~Vr\v{s}nak}
\institute{Hvar Observatory, Faculty of Geodesy, University of Zagreb,\\
 Ka\v{c}i\'ceva 26, HR-10000 Zagreb, Croatia}
\author{M.~Temmer}
\author{A.~Hanslmeier}
\institute{Institute for Geophysics, Astrophysics and Meteorology,
University of Graz, Universit\"atsplatz 5, A-8010 Graz, Austria}

\date{\vspace*{0.4cm} \centering (Received 21 March 2002; accepted 13 April 2002) \vspace*{0.6cm}}
\begin{abstract}
The timing of 503~solar flares observed simultaneously in hard \mbox{X-rays}, soft
X-rays and H$\alpha$ is analyzed. We investigated the start and the peak time
differences in different wavelengths, as well as the differences between the
end of the hard X-ray emission and the maximum of the soft \mbox{X-ray} and
H$\alpha$ emission. In more than 90\% of the analyzed events, a thermal
preheating seen in soft X-rays is present prior to the impulsive flare phase.
On average, the soft X-ray emission starts 3~min before the hard X-ray and the
H$\alpha$~emission. No correlation between the duration of the preheating
phase and the importance of the subsequent flare is found. Furthermore, the
duration of the preheating phase does not differ for impulsive and gradual
flares. For at least half of the events, the end of the nonthermal emission
coincides well with the maximum of the thermal emission, consistent with the
beam-driven evaporation model. On the other hand, for $\sim$25\% of the events
there is strong evidence for prolonged evaporation beyond the end of the hard
\mbox{X-rays}. For these events, the presence of an additional energy transport
mechanism, most probably thermal conduction, seems to play an important role.
\end{abstract}
\end{opening}
%
%\maketitle

\section{Introduction}

In this paper we investigate the timing behavior of solar flares,
observed simultaneously in hard X-ray (HXR), soft X-ray (SXR) and
H$\alpha$ emission. The main items we address are: a) the flare
onset in different wavelengths, b) the timing of different flare
emissions with respect to the electron-heated chromospheric
evaporation model. Based on a sample of 503~solar flares observed
simultaneously in HXR, SXR and H$\alpha$, we aim to determine how
common is the preheating prior to the impulsive phase and is it
different in different types of events. Furthermore, we will
investigate whether the electron beam-driven evaporation model is
consistent with the majority of solar flares, considering the
predicted coincidence between the end of the nonthermal (HXR) and
the maximum of the thermal (SXR and H$\alpha$) flare emission.

Electron beam-driven evaporation is usually supposed to be a dominant energy
transport mechanism during solar flares. According to the thick-target model
(Brown, 1971), the HXR emission is electron-ion bremsstrahlung produced by
electron beams encountering the dense layers of the lower corona, the transition
region, and the chromosphere. The model assumes that only a small fraction of the
energy of the nonthermal electrons is lost through radiation;
most of the energy is transferred to heating of the ambient plasma.
Due to the rapid deposition of energy by the electron beams,
the energy cannot be radiated away sufficiently fast. Thus, a
strong pressure imbalance develops, and the heated plasma explosively
expands up into the corona in a process known as chromospheric evaporation
(Antonucci, Gabriel and Dennis, 1984; Fisher, Canfield and McClymont, 1985;
Antonucci {\it et al.}, 1999). The hot dense plasma that has been convected
into the corona gives rise to enhanced soft SXR emission via thermal bremsstrahlung.
Thus, the model predicts that the hard X-ray emission is directly related to the
flux of the accelerated electrons, whereas the soft X-ray emission is
related to the accumulated energy deposited by the same nonthermal electron population
up to a given time.

However, the model is questioned by several authors. E.g., Simnett (1986)
and Plunkett and Simnett (1994) proposed that protons accelerated at the
energy release site, not electrons, are the primary energy carrier in
solar flares  (see also the review by Simnett, 1995). Another controversial
issue is the role of thermal conduction versus electron beams (e.g.,
Doschek {\it et al.}, 1989). Furthermore, it has been questioned (e.g.,
Feldman, 1990) whether chromospheric evaporation is a ``real'' phenomenon at
all (see the reviews by Doschek {\it et al.}, 1989; Antonucci {\it et al.}, 1999).

From high time resolution observations it is known that during the impulsive phase,
the fast time structures seen in H$\alpha$ are correlated with the hard
X-rays and microwaves (e.g., K\"ampfer and Magun, 1983; Kurokawa, Takakura
and Ohki, 1988; W\"ulser and Marti, 1989; Trottet {\it et al.}, 2000). This
suggests that nonthermal particle beams directly heat the chromospheric plasma,
giving rise to the impulsive H$\alpha$ emission. Numerical simulations of the
chromospheric response to pulse beam heating on time scales of less than 1~s
have been performed by Heinzel (1991). On the other hand, the H$\alpha$ emission
during the main phase of a flare is likely due to heating of the chromosphere by
thermal conduction from the hot SXR emitting plasma in the flare loop (e.g.,
Phillips, 1991). Veronig {\it et al.} (2001) have shown that there is a distinct
correlation between the SXR flux, $F_{\rm SXR}$, and the H$\alpha$ area,
$A_{\rm H\alpha}$, at the time of the flare maximum, close to the relation:
$F_{\rm SXR} \propto (A_{\rm H\alpha})^{3/2}$. This means that the measured
H$\alpha$~area can be understood as an intersection at chromospheric level
of the volume of evaporated plasma responsible for the enhanced SXR emission.

Under the assumption that
exclusively accelerated electrons contribute to the evaporation and that the
cooling time of the plasma is significantly longer than the impulsive HXR
emission (see also Dennis, 1991), it is expected that the SXR as well as
the H$\alpha$ emission do not further increase after the HXR emission, i.e.
the electron input, has stopped. In a previous paper (Veronig {\it et al.}, 2002a),
the timing of the SXR peak emission relative to the end of the HXR emission,
has been investigated. In this study we additionally include H$\alpha$
measurements as a further indicator of thermal flare emission. Considering
H$\alpha$ observations complementary to soft X-rays is of particular interest,
since, as shown by McTiernan, Fisher and Li (1999), the temporal behavior
of the SXR emission depends on the temperature response of the SXR detector used.

There are several papers that investigate the start of soft X-rays relative
to hard X-rays. Kahler (1979) did not find systematic brightenings in
soft X-rays before the impulsive flare phase. On the other hand, Machado,
Orwig and Antonucci (1986) and Schmahl {\it et al.} (1989) reported frequent
strong SXR emission before the impulsive phase. These authors have
shown that, on average, the SXR emission precedes the onset of the HXR emission
by $\sim$2~min. The fact that a gradual rise of SXR emission is present before
the onset of the hard X-rays suggests a thermal origin of the first phase of a
flare (e.g., \v{S}vestka, 1976; Schmahl {\it et al.}, 1989). Such a gradual
heating before the impulsive particle acceleration is supposed to be
related to the re-arrangement of the magnetic fields preceding a flare.
Furthermore, this initial phase may also determine the subsequent impulsive
phase of the flare. For example, Emslie, Li and Mariska (1992) have shown that
preheating of the flare atmosphere influences the subsequent evaporation process.
We stress that this initial phase of the flare observed in soft X-rays has to
be discriminated from SXR precursors, which occur several tens of minutes before
the actual flare and not necessarily at the flare site. Furthermore, there
is a distinct fall of intensity between the precursor event and the flare
itself (e.g., Tappin 1991). In this paper we do not concentrate
on SXR precursors, whose existence is still a controversial issue, but refer
to the papers by Webb (1983), Tappin (1991), F\'{a}rn\'{\i}k, Hudson and Watanabe (1996),
and F\'{a}rn\'{\i}k and Savy (1998).

Statistical studies of the timing behavior of solar flares
observed at different wavelengths have been presented in several
papers. However, in general only the emissions at two wavelengths
are compared. Investigations on the timing of the SXR and
H$\alpha$ emission during a flare have been carried out by Thomas
and Teske (1971), Datlowe, Hudson and Peterson  (1974), Falciani
{\it et al.} (1977), Zirin {\it et al.} (1981), Verma and Pande
(1985a) and Veronig {\it et al.} (2002b). The reported results are
quite contradictory. There is neither a consensus whether the SXR
emission starts before the H$\alpha$ emission or vice versa, nor
in which succession the peaks occur. These conflicting results are
presumably related to the fact that apart from the studies by
Thomas and Teske (1971), Datlowe, Hudson and Peterson (1974) and
Veronig {\it et al.} (2002b), rather small data sets have been
used for the analysis, which may cause large statistical errors.
A statistical study of the timing of the HXR emission relative to
the H$\alpha$ emission has been performed by Verma and Pande
(1985b), finding that most impulsive flares produce HXR emission
up to 1~min before and up to 2~min after the start of the
H$\alpha$ emission. For previous studies of the relative timing of
the SXR and HXR flare emission we refer to Machado, Orwig and
Antonucci (1986) and Schmahl {\it et al.} (1989), and references
therein.

\section{Data Selection}

We utilize the SXR data from the Geostationary Operational
Environmental Satellites (GOES), the HXR data observed by the
Burst and Transient Source Experiment (BATSE) aboard the
Compton Gamma-Ray Observatory (CGRO) and the H$\alpha$ flare
data collected by the {\it Solar Geophysical Data} (SGD).
The X-ray sensor aboard GOES consists of two ion chamber
detectors, which provide whole-sun X-ray flux measurements in the
0.5--4 and 1--8~{\AA} wavelength bands. The GOES X-ray sensor is
described in Donelly and Unzicker (1974) and Garcia (1994). BATSE
is a whole-sky HXR flux monitor made up of eight large-area
wide-field detectors. From each detector there are hard X-ray
measurements in four energy channels, 25--50, 50--100, 100--300
and $>$300~keV. A characteristics of the BATSE instrument and its
capabilities for solar flare studies can be found in Fishman
{\it et al.} (1989, 1992) and Schwartz {\it et al.} (1992).

For the analysis, the 1-min averaged GOES SXR data measured in the
1--8~{\AA} channel, the HXR data collected in the BATSE Solar Flare
Catalog, archived in the Solar Data Analysis Center at NASA/Goddard
Space Flight Center, and the H$\alpha$ flare reports from the SGD
for the period January 1997 to June 2000 are used. In the BATSE
Flare Catalog, the start, maximum and end times of an HXR event
are reported with an accuracy of 1~s, whereas the GOES SXR and the
H$\alpha$ flare reports in the SGD have a 1~min precision. Thus,
we cannot expect to obtain any reliable time difference with an
accuracy $\lesssim$1~min. For the comparison of the HXR
start/peak/end times with the characteristic times of the SXR and
H$\alpha$ emission we also round off the HXR times to minutes.

Some points regarding the used data should be
stressed. Various observatories around the world contribute to the H$\alpha$
flare data collected by the SGD. In recent years, most (or all) of
the observatories perform regular observations on a time cadence of
1~min or better (Helen Coffey, private communication).
However, there may be still a spread in the reported H$\alpha$ flare
times among the different observatories (due to different instruments,
seeing conditions, etc.) In the selection of the H$\alpha$ flares
from the SGD data collection, we excluded all events in which the start
or maximum time was annotated as uncertain or in which the times
reported by different observatories differed for more than two minutes.
For the X-ray emissions, it has to be noted that the peak times in both
soft and hard X-rays are precisely known. However, the situation is less
clear for the start times. Since the soft X-rays are stronger than the
hard X-rays by a few magnitudes, the sensitivity thresholds play an
important role. Due to the lower relative sensitivity in hard X-rays,
the start of the HXR flare is detected later than it actually occurs.
In particular, in short and weak HXR events, the start time
may be subject to substantial uncertainties.

The identification of corresponding HXR/SXR/H$\alpha$ events is
based on temporal coincidence. To be
attributed as corresponding events, we demand that at all three
wavelengths the emission starts within a time window of 10~min.
Furthermore, in order to avoid any incidental assignment, we applied
the following refinements. Events that overlap in time with any
other event observed at the same wavelength are excluded from the
analysis. Also events for which a multiple assignment to flares
observed at any of the other two wavelengths is possible, e.g.
one SXR event can be assigned to two HXR events, are excluded.
Applying these selection criteria, for the considered period we
obtained 503~flares that are observed simultaneously in soft X-rays,
hard X-rays and H$\alpha$.

\section{Results}

For each event we determined the difference of the start times as
well as peak times in HXR, SXR and H$\alpha$ emission. Moreover,
we derived the difference between the HXR end time and the SXR and
H$\alpha$ peak time. The differences are determined in absolute
values (given in minutes), denoted here as $\Delta t_{\rm abs}$, as well
as normalized to the duration~$D$ of the respective HXR event, i.e.
\begin{equation}
\Delta t_{\rm norm} = \frac{\Delta t_{\rm abs}}{D} \, .
\end{equation}
The normalized time differences, $\Delta t_{\rm norm}$, are of
particular interest when the timing behavior of long-duration
flares is considered. Such events may show distinct time
differences in absolute values; however, the time difference may
be rather small compared to the overall endurance of the event.
Moreover, since the importance of a flare and its duration are
correlated (e.g., Crosby {\it et al.}, 1998; Veronig {\it et al.}, 2002a,c)
considering only absolute time differences may introduce a bias
towards intense events (for further discussion see Veronig {\it et al.},
2002a). From the present data set we derive a median duration of
2.0~min for the BATSE HXR bursts.

\begin{figure}
\centering
\resizebox{\hsize}{!}{\includegraphics{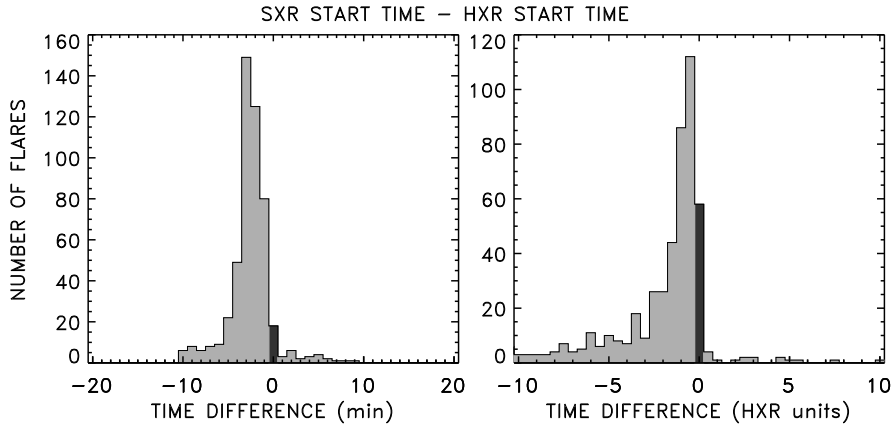}}
    \caption{Histogram of the SXR--HXR start time differences,
    given in absolute values (left panel) and normalized to
    the HXR event duration (right panel). }
\end{figure}

Figure~1 shows the histogram of the start time differences of the SXR
and the HXR emission derived from the total set of 503~events, in
absolute values (left panel) and normalized to the HXR event
duration (right panel). For the absolute time differences a bin
size of 1~min, and for the normalized time differences a bin size of
0.5 units of the HXR duration is applied. (Note that Figures~2--8 are
constructed analogously to Figure~1.) From the histograms it is clearly
evident that in most events the SXR emission starts before the HXR
emission. Only in 5\% of the considered flares, the SXR emission
starts after the HXR emission. The distribution of the absolute time
differences, $\Delta t_{\rm abs}$, has its mode at $\Delta t_{\rm abs} = -3$~min.
For the normalized time differences, $\Delta t_{\rm norm}$, the overall
situation is quite similar. However, due to the weighting with the HXR
event duration there are more events that belong to the bin
$\Delta t_{\rm norm} = 0$, indicative for a time difference
less than 0.25 times the HXR event duration. The mode of the distribution is
located at $\Delta t_{\rm norm} = -0.5$~HXR units.

In Figure~2, the respective distributions of the SXR--H$\alpha$ start time differences
are plotted. The general behavior is quite similar to
those of the SXR--HXR start times. For the bulk of flares, the SXR
event starts before the H$\alpha$ event; only in 12\% of the
cases the SXR emission starts after the H$\alpha$ emission. The mode of the
distribution of absolute time differences is located
$\Delta t_{\rm abs} = -3$~min; for the normalized time differences it lies
at $\Delta t_{\rm norm} = -0.5$~HXR units.

\begin{figure}
\centering
\resizebox{\hsize}{!}{\includegraphics{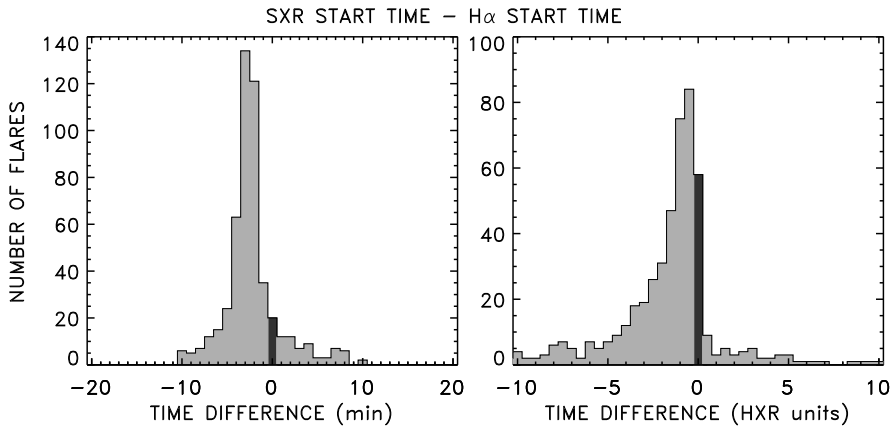}}
    \caption{Histogram of the SXR--H$\alpha$ start time differences,
    given in absolute values (left panel) and normalized to
    the HXR event duration (right panel). }
\end{figure}

Figure~3 shows the histograms of the H$\alpha$--HXR start time differences.
The distributions of the absolute as well as normalized time differences have a
distinct peak at zero, indicating that within the given precision, the
H$\alpha$ and HXR emissions preferentially start simultaneously.
$27\%$ of the events start within the same minute, and
$60\%$ start within $|\Delta t_{\rm abs}| \le 1$~min. The normalized
time differences are even more concentrated around zero: 37\% of the
events are covered by the bin $\Delta t_{\rm norm} = 0$. Furthermore,
we note that the total number of events with negative time differences is
similar to those with positive time differences (see also Table~I).

\begin{figure}
\centering
\resizebox{\hsize}{!}{\includegraphics{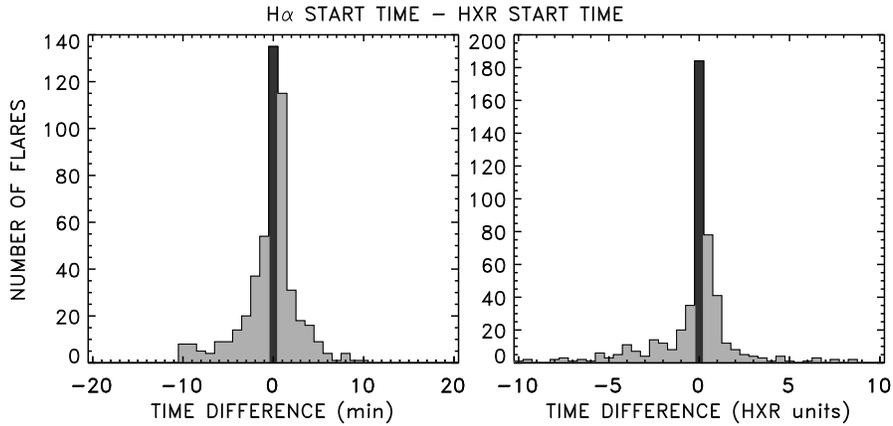}}
    \caption{Histogram of the H$\alpha$--HXR start time differences,
    given in absolute values (left panel) and normalized to
    the HXR event duration (right panel). }
\end{figure}

Comparing the distributions of the absolute time differences of the
SXR, HXR and H$\alpha$ start times (Figures~1--3, left panels), it can be seen
that by the used criterion we certainly have missed flares, in which
the SXR emission starts more than 10~min prior to the HXR or H$\alpha$ emission.
On the other hand, there might be only a negligible number of events, in which
the SXR emission starts more than 10~min later than the HXR or H$\alpha$
emission. In general, the distributions in \mbox{Figures~1--3}
confirm the reasonableness of the used 10~min start time window for the
identification of corresponding events, since the shape of the histograms
indicates that relatively few events lie outside this range.

\begin{figure}
\centering
\resizebox{\hsize}{!}{\includegraphics{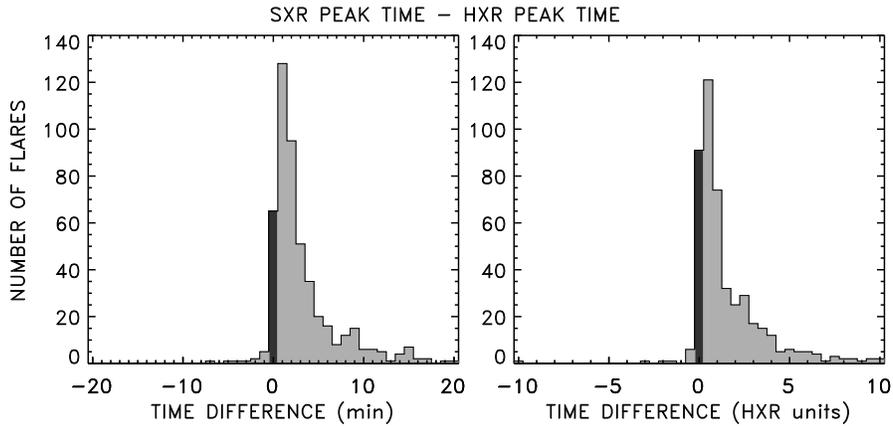}}
    \caption{Histogram of the SXR--HXR peak time differences, given in
    absolute values (left panel) and normalized to
    the HXR event duration (right panel). }
\end{figure}

\begin{figure}
\centering
\resizebox{\hsize}{!}{\includegraphics{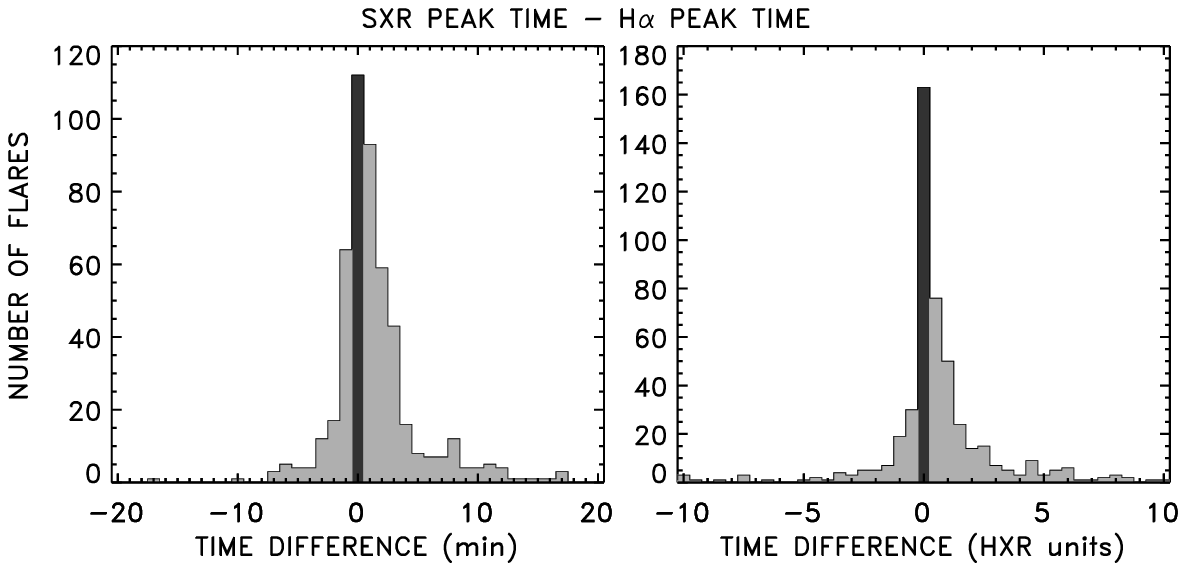}}
    \caption{Histogram of the SXR--H$\alpha$ peak time differences,
    given in absolute values (left panel) and normalized to
    the HXR event duration (right panel). }
\end{figure}

\begin{figure}
\centering
\resizebox{\hsize}{!}{\includegraphics{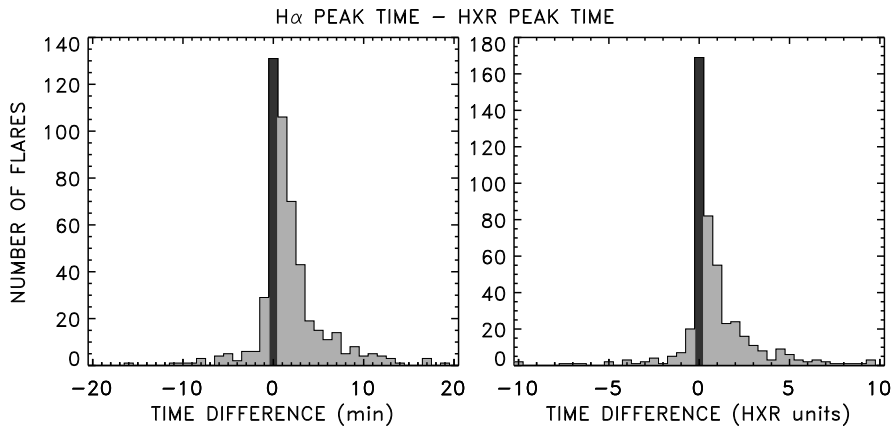}}
    \caption{Histogram of the H$\alpha$--HXR peak time differences,
    given in absolute values (left panel) and normalized to
    the HXR event duration (right panel). }
\end{figure}

In Figures~4, 5 and 6 the distributions of the peak time differences
between the SXR and HXR, the SXR and H$\alpha$, and the H$\alpha$ and
HXR emission are represented, respectively. Figure~4 shows that the
soft X-rays predominantly peak after the hard X-rays, with the mode at
$\Delta t_{\rm abs} = +1$~min and $\Delta t_{\rm norm} = +0.5$~HXR units,
respectively. Only in 2\% of the events, the SXR maximum takes place
before the HXR maximum. In contrast to that, the SXR and the H$\alpha$
emission peak preferentially simultaneously. The absolute as well as
the normalized distributions have its mode at zero. Yet, the distribution
is asymmetric: there are more events, for which the SXR emission
peaks after the H$\alpha$ emission than {\it vice versa} (cf. Figure~5).
From Figure~6 it can be inferred that also the histograms of the
time differences of the H$\alpha$ and HXR peak emission have its mode at
zero, but there exists a clear tendency that the H$\alpha$ emission peaks
after the HXR emission. Only in 12\% of the events, the H$\alpha$ maximum
takes place before the HXR peak.

In Figure~7, the distributions of the differences of the SXR peak time
and the HXR end time are plotted. The absolute and the
normalized time differences have its maximum at zero. For 22\% of the
events the SXR peak and the HXR end take place within the same
minute, 53\% lie within $|\Delta t_{\rm abs}| \le 1$~min. Yet, there
are more events, for which the SXR maximum occurs after the HXR end
than {\it vice versa}. This behavior is in particular evident from the distribution
of the normalized time differences. Almost all events with
negative time differences belong to the bin $\Delta t_{\rm norm} = -0.5$,
whereas the events with positive time difference distribute over a
broad range of $\Delta t_{\rm norm}$. This indicates that many of the
flares located on the left hand side of the distribution, embracing events in
which the SXR emission is already decreasing while there is still
HXR emission detectable, are of long duration.
For the events with negative time difference, we obtain a median HXR duration
of $\bar{D} = 4.4$~min, whereas for the events with positive time difference
we find $\bar{D} = 1.2$~min.

\begin{figure}
\centering
\resizebox{\hsize}{!}{\includegraphics{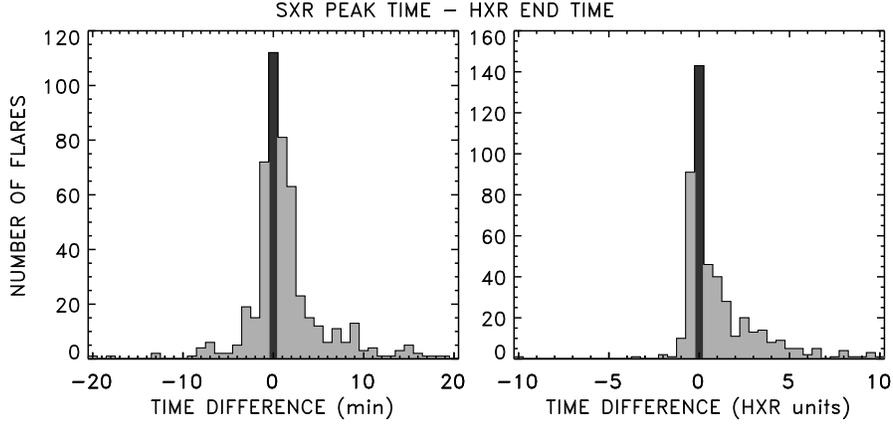}}
    \caption{Histogram of the differences of the SXR peak times and the HXR
    end times, given in absolute values (left panel) and normalized to
    the HXR event duration (right panel). }
\end{figure}
\begin{figure}
\centering
\resizebox{\hsize}{!}{\includegraphics{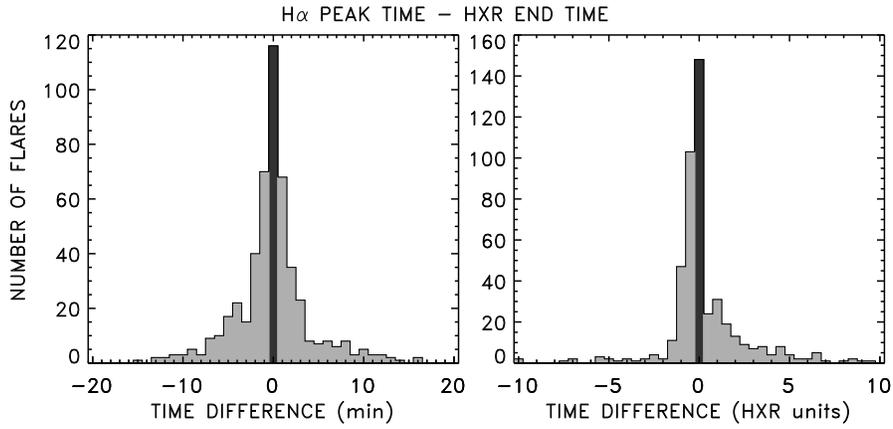}}
    \caption{Histogram of the differences of the H$\alpha$ peak times and the HXR
    end times, given in absolute values (left panel) and normalized to
    the HXR event duration (right panel). }
\end{figure}

In principle, it is expected that there are no events with
\mbox{$\Delta t_{\rm norm} < -1$} HXR unit, since this range
indicates that the maximum of the SXR emission occurs before the
start of the HXR emission. However, as can be seen from Figure~7, a
few such events exist. There are two possible explanations for that:
a)~the respective HXR and SXR events are not causally related (e.g.,
take place at different sites), b)~the determined HXR burst
duration is too short due to the sensitivity limits of the HXR
detectors. In particular, in weak and short HXR events, a considerable
part of the HXR burst duration will be missed, if the emission is not
distinctly higher than the noise level.

Figure~8 shows the distributions of the differences of the H$\alpha$ peak time
and the HXR end time. The histogram of the absolute time differences
is strongly peaked at $\Delta t_{\rm abs}$ = 0~min. 23\% of the
events are covered by $\Delta t_{\rm abs}$ = 0~min, 51\% lie within
$|\Delta t_{\rm abs}| \le 1$~min. A similar behavior shows up for the
normalized time differences. Similar to Figure~7, but less pronounced, the range with
negative time differences reveals a tendency to cover long-duration flares.
For the events with negative time difference we obtain $\bar{D} = 3.6$~min,
for those with positive time differences $\bar{D} = 1.1$~min.

In Table~I and~II, a summary of the histograms represented in Figures~1--8
is given. Table~I specifies the percentage of events, for which the absolute
time differences, $\Delta t_{\rm abs}$, are negative, positive or
zero (i.e. the respective emissions take place within the same minute).
In Table~II, we list the mode and the median of the absolute and the
normalized time differences, characterizing the maximum and the center of
the distributions, respectively.

\begin{table}
\caption{For the various time differences derived (cf. Figures~1--8), we list
the percentage of events, for which the time difference is negative, positive or zero.}
\begin{tabular}{llrrr} \hline
Fig. & Time difference & \multicolumn{3}{c}{Number of events (\%)} \\
&  & \multicolumn{1}{c}{$\Delta t_{\rm abs} < 0$} & \multicolumn{1}{c}{$\Delta t_{\rm abs} > 0$}
 & \multicolumn{1}{c}{$\Delta t_{\rm abs} = 0$} \\ \hline
1 & SXR start -- HXR start         & 91.8~~ & 4.6~~  & 3.6~~ \\
2 & SXR start -- H$\alpha$ start   & 83.9~~ & 12.1~~ & 4.0~~ \\
3 & H$\alpha$ start -- HXR start   & 33.4~~ & 39.8~~ & 26.8~~ \\
4 & SXR peak -- HXR peak           & 2.2~~  & 84.9~~ & 12.9~~ \\
5 & SXR peak -- H$\alpha$ peak     & 22.2~~ & 55.5~~ & 22.3~~ \\
6 & H$\alpha$ peak -- HXR peak     & 11.7~~ & 62.2~~ & 26.1~~ \\
7 & SXR peak -- HXR end            & 25.8~~ & 51.9~~ & 22.3~~ \\
8 & H$\alpha$ peak -- HXR end      & 40.5~~ & 36.4~~ & 23.1~~ \\ \hline
\end{tabular}
\end{table}

\begin{table}
\caption{For the various time differences derived (cf. Figures~1--8), we list
the median and the mode of the absolute and normalized time differences, respectively.}
\begin{tabular}{llrrrr} \hline
Fig. & \multicolumn{1}{l}{Time difference} & \multicolumn{2}{c}{$\Delta t_{\rm abs}$ (min)} & \multicolumn{2}{c}{$\Delta t_{\rm norm}$ (unit)} \\
&   & \multicolumn{1}{c}{Median}  &  \multicolumn{1}{c}{Mode} & \multicolumn{1}{c}{Median}  &  \multicolumn{1}{c}{Mode}   \\ \hline
1 & SXR start -- HXR start         & $-3.0~~$ & $-3.0~$ & $-1.1~~$  & $-0.5~$  \\
2 & SXR start -- H$\alpha$ start   & $-3.0~~$ & $-3.0~$ & $-1.1~~$  & $-0.5~$ \\
3 & H$\alpha$ start -- HXR start   & $ 0.0~~$ & $ 0.0~$ & $ 0.0~~$  & $ 0.0~$ \\
4 & SXR peak -- HXR peak           & $+2.0~~$ & $+1.0~$ & $+0.9~~$  & $+0.5~$ \\
5 & SXR peak -- H$\alpha$ peak     & $+1.0~~$ & $ 0.0~$ & $+0.2~~$  & $ 0.0~$ \\
6 & H$\alpha$ peak -- HXR peak     & $+1.0~~$ & $ 0.0~$ & $+0.5~~$  & $ 0.0~$ \\
7 & SXR peak -- HXR end            & $+1.0~~$ & $ 0.0~$ & $+0.3~~$  & $ 0.0~$ \\
8 & H$\alpha$ peak -- HXR end      & $ 0.0~~$ & $ 0.0~$ & $ 0.0~~$  & $ 0.0~$ \\ \hline
\end{tabular}
\end{table}

\section{Discussion}

\subsection{Flare onset}

We find that in more than 90\% of the analyzed events the SXR
emission starts before the HXR emission for at least 1~min.
About 80\% of the analyzed events lie within the range:
$ -4\; {\rm min} \le \Delta t_{\rm abs} \le -1\;{\rm min}$. It has
to be noted that the cut-off on the left hand side of the histogram of
the SXR--HXR start time differences (Figure~1, left panel)
strongly suggests that there are also events, for which the SXR emission starts even more
than 10~min before the HXR emission, not covered by our analysis.
Moreover, we obtain that in 84\% the SXR emission starts
before the H$\alpha$ emission. Only in 5\% (12\%) of the sample, the SXR
emission starts later than the HXR (H$\alpha$) emission (cf. Table~1).
On the other hand, the H$\alpha$ and the HXR emission start
preferentially simultaneously. The distribution shows a very sharp
peak at zero, which is in particular evident for the normalized time differences,
and a quite symmetric behavior for positive and negative time
differences (cf. Figure~3). This provides evidence that the onset of
the H$\alpha$ emission is related to the impulsive phase of particle
acceleration, in which the chromosphere is directly heated by electron
bombardment. This finding coincides well with high time resolution
observations, from which strong temporal correlations between the
hard X-ray and H$\alpha$ fine structures during the impulsive
flare phase are reported (e.g., Kurokawa, Takakura and Ohki, 1988;
W\"ulser and Marti, 1989; Trottet {\it et al.}, 2000).

In principle, an enhanced thermal emission preceding the onset of the hard
X-rays may be indicative of a thermal preheating phase prior to the impulsive
electron acceleration, or it can be caused by the sensitivity threshold
of the hard X-ray detectors (e.g., Dennis, 1988). We find that, on average, the
SXR emission precedes the hard \mbox{X-rays} by 3~min. This corresponds to more
than 1~unit of the HXR event duration for the normalized time differences (cf. Table~2).
For about one third of the events, it is even more than 2~times the HXR duration,
which can hardly be explained by the sensitivity limits of the HXR detectors.
On average, we find slightly longer preceding SXR emissions than Machado,
Orwig and Antonucci (1986) and Schmahl {\it et al.} (1989), who reported an average
value of 2~min, although the BATSE HXR detectors are far more sensitive than
previously used instruments (Schwartz {\it et al.}, 1992). So, relating
the preceding soft X-ray emission to the sensitivity threshold of the HXR
detectors, we would expect to obtain a shorter duration for this phase
than previous authors. Furthermore, we find that in most of the events the
soft X-ray emission precedes not only the HXR but also the H$\alpha$ emission.
All these findings suggest that the preceding SXR emission can be interpreted
in terms of a thermal preheating of the flare atmosphere prior to the
impulsive particle acceleration, and that such preheating occurs in almost
all flares.

No correlation between the duration of the preheating phase and the importance
of the subsequent flare is found. The cross-correlation coefficients,~$r$,
determined between the duration of the preheating phase, $\Delta t_{\rm abs}$, and
the HXR peak emission, the SXR peak emission and the H$\alpha$ area
yield $|r| \lesssim 0.1$. This means that preheating occurs in the same manner
in weak as well as intense events. Moreover, there is neither a
significant correlation between $\Delta t_{\rm abs}$ and the duration of the
HXR event nor between $\Delta t_{\rm abs}$ and the HXR peak flux divided by
the HXR rise time. This outcome suggests that there is no distinct
difference in impulsive and gradual flares.

Current sheet models (e.g., Heyvaerts, Priest and Rust, 1977) of solar
flares predict the existence of a preheating phase (see also the
review by Gaizauskas, 1989). Li, Pallavicini and Cheng (1987) have
discussed the preheating of the flare atmosphere caused by emerging
magnetic flux, and the formation of a current sheet at the interface
between the newly emerging flux tube and the overlying magnetic region.
The authors obtain a spectrum of solutions depending on the rate of
flux emergence. This can explain gradual enhancements of the SXR emission
without any impulsive electron acceleration\footnote{It has to be noted that this
case, SXR burst without accompanying electron acceleration, i.e.\ without
HXR emission, is a priori excluded in our analysis, studying corresponding
SXR/HXR/H$\alpha$ events.}, as well as the gradual preheating
stage seen in soft X-rays before an impulsive flare phase. Vr\v{s}nak (1989)
has shown that in fact a hot turbulent current sheet characterized by an increasing
merging velocity can very well explain the SXR emission during the early phases
of solar flares regardless on the details of the magnetic field configuration.

Emslie, Li and Mariska (1992) have shown that preheating influences the subsequent
chromospheric evaporation process, insofar as it causes smaller flow speeds of the evaporated
material. The main argument is the following: If the flare plasma is preheated,
the initial coronal density is enhanced. This leads to an increase of the
column depth of the corona, and the accelerated electrons loose their energy
predominantly in coronal layers. Thus, chromospheric evaporation will not be
provoked. Moreover, the large inertia of the corona will suppress any upward
velocity. Taking into account these effects of preheating, the simulations
of the electron-heated flare model reproduced the evaporation flow speeds
known from spectroscopic observations (Emslie, Li and Mariska, 1992),
whereas not taking into account a preheated flare atmosphere, the derived flow
speeds were too high compared to observations (Li, Emslie and Mariska, 1989).

\subsection{Relation between the peak and the end of the
thermal and nonthermal flare emissions}

For the timing of the emission peaks (cf.~Figures 4--6), we obtain that in
most flares the SXR and H$\alpha$ emission reach their maximum after the
HXR emission. Only in 2\% (12\%) of the considered events, the SXR (H$\alpha$)
emission peaks prior to the hard X-rays. The fact that the SXR and H$\alpha$ emission
predominantly peak after the HXR emission, can be considered as a necessary condition
for the electron beam-driven evaporation model (e.g. Benz, 1995), as the model assumes
that the thermal flare emission is caused by thermalization of the
same nonthermal electrons that are emitting in hard X-rays. The SXR and
the H$\alpha$ emission peak preferentially simultaneously. Yet, the
distribution is not symmetric but reveals a tendency for the soft X-ray
peak taking place after the H$\alpha$ maximum (cf. Table~1). This
difference may be interpreted as an indication that the chromospheric plasma
is cooling faster than the coronal plasma.

The most direct test for the electron beam-driven chromospheric
evaporation model is provided by Figures~7 and~8, representing the
distributions of the differences of the SXR (H$\alpha$) peak time
and the HXR end time. If the electron-heated evaporation model
holds, then the thermal emissions should not further increase
after the HXR emission, i.e. the energy supply by nonthermal
electrons, has stopped. Indeed, both distributions have its mode at
zero, and more than half of the events lie within $|\Delta t_{\rm abs}| \le 1$~min.
However, there is also a significant fraction that violates
the predicted timing behavior. In the following, we discuss
separately the two different cases of deviation from $\Delta t = 0$,
i.e. negative and positive time differences.

12\% (27\%) of the events have $\Delta t_{\rm abs} < -1$~min, i.e.
the SXR (H$\alpha$) emission already decreases although the HXR
emission is still present for more than 1~min. As it is shown above,
these events are preferentially flares with
long HXR duration, i.e. gradual flares. Thus, the assumption that the
cooling time of the plasma is distinctly longer than the impulsive HXR
emission is not valid any longer. The outcome that in gradual events
the thermal emission peaks before the end of the nonthermal HXR
emission is compatible with the electron beam-driven evaporation model.
Li, Emslie and Mariska (1993) simulated HXR and SXR light curves
from a thick-target electron-heated model, finding that in gradual
flares the maximum of the SXR emission occurs before the end of the HXR
event, caused by the fact that during the decay phase of the HXR flare,
the energy supply by the evaporation-driven density enhancements cannot
overcome the instantaneous cooling of the hot plasma.

36\% (23\%) of the events have $\Delta t_{\rm abs} > 1$~min, i.e. the SXR
(H$\alpha$) emission is still increasing although the HXR emission, i.e.
the energy input by the nonthermal electrons, already stopped more
than 1~min earlier. Such outcome, in principle, suggests that the hot plasma
giving rise to the thermal emission is not heated exclusively by thermalization
of the accelerated electrons, which are responsible for the HXR emission.
However, as argued by McTiernan, Fisher and Li (1999), an extended
heating beyond that due to nonthermal electron beams is not the only
possible explanation for flares, in which the SXR emission is still
increasing although the HXR emission already stopped, but it might be related
to the temperature response of the SXR detector. If the used SXR
detector has also a substantial response to low-temperature
plasma, and the GOES 1--8~{\AA} detector has (James McTiernan, private communication),
then the extended increase of the SXR emission might be due to cooling of
high-temperature flare plasma (for further discussion see McTiernan,
Fisher and Li, 1999).

This may provide an explanation, why we find more events with a prolonged
increase of SXR emission than events with a prolonged increase of H$\alpha$ emission
beyond the end of the hard X-rays (cf. Table~1). However, also in the case of the
H$\alpha$ emission, which is not affected by the above argument, there is
a significant fraction of events (about one third), in which the emission
is still increasing after the hard X-rays have stopped. For 24\% of the events,
the H$\alpha$ as well as the SXR emission of an event is increasing for longer than
1~min or longer than 1~unit of the HXR event duration after the HXR emission
already stopped. This subset is indicative of a strong violation of the timing
behavior predicted from the electron beam-driven evaporation model. Thus, for
a significant fraction there is strong evidence for an additional agent
other than the HXR emitting electrons, contributing to the energy input and prolonging
the evaporation. It is important to note that these events have shorter duration
and smaller HXR fluxes than the average. From the subset with distinct prolonged
thermal emissions we find $\bar{D} = 0.8$~min and
$\bar{F}_{\rm HXR} = 666$~counts~s$^{-1}$\,/\,2000~cm$^2$,
whereas from the complete sample we obtain $\bar{D} = 2.0$~min and
$\bar{F}_{\rm HXR} = 1510$~counts~s$^{-1}$\,/\,2000~cm$^2$.
In Figure~9, we show the distribution of the HXR peak fluxes (panel~{\bf a})
and the HXR event duration (panel {\bf b}). The distributions of the
complete sample are shaded in light grey, the distributions of the
subset characterized by a distinct prolongation of the thermal emissions
in dark grey. From the figure it is evident that this subset contains
almost no events with high HXR fluxes or long HXR duration.

\begin{figure}
\centering
\resizebox{\hsize}{!}{\includegraphics{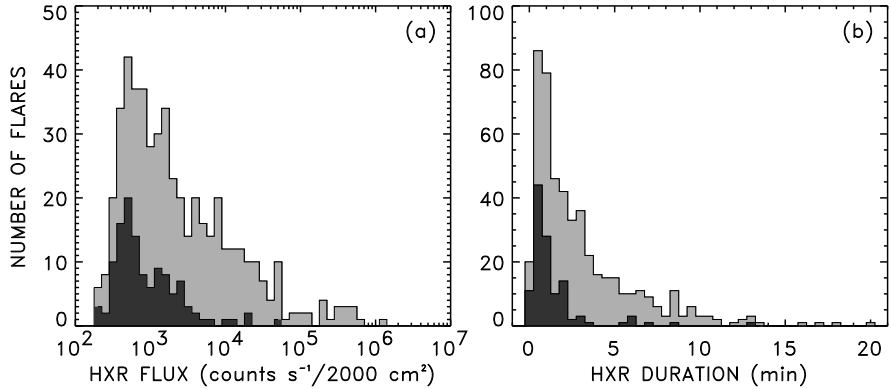}}
    \caption{Panel {\bf a} shows the distribution of the HXR peak fluxes,
    derived from the complete sample ({\it light grey}), and separately from
    the events, in which the increase of the thermal emission is distinctly
    prolonged after the HXR end ({\it dark grey}). In panel~{\bf b}, the
    respective distributions of the HXR burst duration are plotted. }
\end{figure}

\subsection{Interpretation with regard to different energy transport mechanisms}

Simnett (1986) and Plunkett and Simnett (1994) have proposed that
accelerated protons drive the evaporation process, whereas
accelerated electrons play an unimportant role in terms of
the energy transport in solar flares. In the frame of proton
dominated models, the hard \mbox{X-rays} are either produced directly by protons
via inverse bremsstrahlung (e.g., Emslie and Brown, 1985; Heristchi, 1986),
or by electrons, secondary accelerated in the chromosphere by
proton or neutral beams (Simnett and Haines, 1990). Recently,
Karlick{\'y} {\it et al.} (2000) have proposed a model, in which
hard X-rays may be emitted by a neutral beam due to heating of the
electrostatically dragged beam electrons in collision with a neutral
background plasma. As argued in the review by Simnett (1995), models
invoking proton beams can account for several observational
features, which are difficult to be explained in the frame of electron beam
models, as, e.g., preceding soft X-rays, delay of the microwaves with regard
to the hard X-rays, plasma flows prior to the HXR emission
(but see also the criticism, e.g., in Aschwanden, 1996; Emslie {\it et al.}, 1996;
Newton, 1997).

In principle, from the present analysis, finding that for at least
half of the events there is a good agreement between the end of the HXR emission
and the maximum of the thermal SXR and H$\alpha$ emission, we cannot
exclude that proton beams rather than electron beams play the
dominant role in the energy transport. Such temporal coincidence
is expected from the proton beam model in the same way as from the
electron beam model (at least within the given accuracy of 1~min).
Thus, the electron versus proton debate is beyond the capabilities
of this study. However, from energy-dependent timing delays
inferred from high time resolution hard X-ray observations,
Aschwanden (1996) concluded that protons can be ruled out as the
primary energy carrier in solar flares. New insight into the site and
the nature of the energy deposition in solar flares, and the contribution
of high-energy particles to the flare energetics is anticipated by the
recently launched {\it Reuven Ramaty High Energy Solar Spectroscopic
Imager} (RHESSI) mission.

On the other hand, for $\sim$25\% of the analyzed events there is
strong evidence that the thermal emission is significantly prolonged beyond
the end of the HXR emission. This finding is in contradiction to both the
electron and the proton beam-driven evaporation model, and implies that for
a considerable fraction of the analyzed flare sample, there is evidence
for an additional energy transport mechanism other than electron (or proton)
beams. A promising candidate is thermal
conduction, simply due to the fact that the energy release site is strongly
heated (for discussion see Vr\v{s}nak, 1989; Somov, 1992;
McDonald, Harra-Murnion and Culhane, 1999; and references therein).
Furthermore, in several papers observational evidence for
thermal conduction driving the evaporation process is reported.
From high time resolution HXR and H$\alpha$ observations,
K\"ampfer and Magun (1983) and W\"ulser and Marti (1989) found
indications for the occurrence of energy transport by electron beams
at one flare kernel and for conductive energy transport at another
kernel. Similar, using spatially resolved H$\alpha$ measurements
combined with HXR and SXR observations, Kitahara and Kurokawa (1990) inferred
the presence of at least two different energy transport mechanisms, most probably
electron precipitation and thermal conduction. Rust, Simnett and
Smith (1985) report observational evidence for thermal conduction
fronts seen in SXR images. These authors report that the inferred heat
flux accounts for most of energy released in the studied flares.
Furthermore, various observations lend support for a prolonged
chromospheric evaporation driven by thermal conduction
fronts during the decay phase of solar flares (e.g., Zarro and Lemen 1988;
Schmieder {\it et al.} 1990; Czaykowska, Alexander, and de Pontieu, 2001).

\section{Conclusions}

We found in almost all of 503~studied flares ($\sim$90\%) a thermal
preheating of the flare atmosphere, seen in soft X-rays prior to
the impulsive particle acceleration. On average, the SXR emission starts
$\sim$3~min before the HXR and the H$\alpha$ emission. The duration of the
preheating phase is not related to the importance of the subsequent flare.
Moreover, there is no evidence that the duration of the preheating phase
differs for impulsive and gradual flares. The H$\alpha$ and the HXR emission
start preferentially simultaneously, indicating that the onset of the
H$\alpha$~emission is related to the impulsive phase of particle acceleration.

The thermal (SXR and H$\alpha$) emissions predominantly peak after the
nonthermal (HXR) emission. This outcome provides a necessary condition
for the electron-heated chromospheric evaporation model, in which
the thermal flare emission is caused by thermalization of the
same nonthermal electrons that are emitting in hard X-rays. For more
than half of the events, the end of the nonthermal emission coincides well
with the maximum of the thermal emission ($\Delta t \le 1$~min), as predicted
from the beam-driven evaporation model. However, for  $\sim$25\% of the
events, there is strong evidence for a prolonged evaporation beyond the end
of the nonthermal emission. On average, these events are characterized by a
weak and short HXR emission. The extended thermal emission beyond the hard
\mbox{X-rays} suggests the presence of an additional energy transport
mechanism from the energy release site other than particle beams,
most probably thermal conduction. Events, in which the thermal emission is
found to peak prior to the end of the hard \mbox{X-rays}, are preferentially
of long duration. This effect can be explained, within the electron-heated
evaporation model, by instantaneous cooling of the plasma that
dominates over the energy supply by evaporated material during the decay
phase of long-duration events.

\acknowledgements
The authors thank Brian Dennis, Richard A. Schwartz and A. Kimberley
Tolbert from the BATSE team for making available the solar flare data,
as well as Helen Coffey from NGDC to provide the SXR data. We
thank also the referee Franti\v{s}ek F\'arn\'{\i}k for clarifying
comments, which made the paper more precise. A.V., M.T. and A.H.
gratefully acknowledge the Austrian {\em Fonds zur F\"orderung der
wissenschaftlichen Forschung} (FWF grants P13653-PHY and P15344-PHY)
for supporting this project. B.V. acknowledges the University of Graz
for financial support and is grateful to the colleagues from the
Institute for Geophysics, Astrophysics and Meteorology for their hospitality.

\end{article}
\end{document}